\documentclass{article}
\usepackage{graphicx}
\usepackage{latexsym}

\begin{document}
\begin{flushright}
Accepted for publication in Physical Review D\\
UM-P 014-2001\\
RCHEP 001-2001
\end{flushright}
\begin{center}
{\Large\bf TeV-scale electron Compton scattering in the Randall-Sundrum
scenario.}\\
\hspace{10pt}\\
S.R.Choudhury\footnote{src@ducos.ernet.in}\\
{\em International Centre for Theoretical Physics, Trieste, Italy}\\
{\em and School of Physics, University of Melbourne,}\\
{\em Parkville, Victoria 3108, Australia}\\
A.S.Cornell\footnote{a.cornell@tauon.ph.unimelb.edu.au}, 
and G.C.Joshi\footnote{joshi@physics.unimelb.edu.au}\\
{\em School of Physics, University of Melbourne,}\\
{\em Parkville, Victoria 3108, Australia}\\
\end{center}
\hspace{10pt}\\
\begin{abstract}
\indent The spin-2 graviton excitations in the Randall-Sundrum gravity model
provides a $t$-channel contribution to electron Compton scattering which
competes favourably with the standard QED contributions.  The
phenomenological implications of these contributions to the unpolarized
and polarized cross-sections are evaluated.
\end{abstract}
\begin{figure}
\begin{center}
\includegraphics[angle=0,width=10cm]{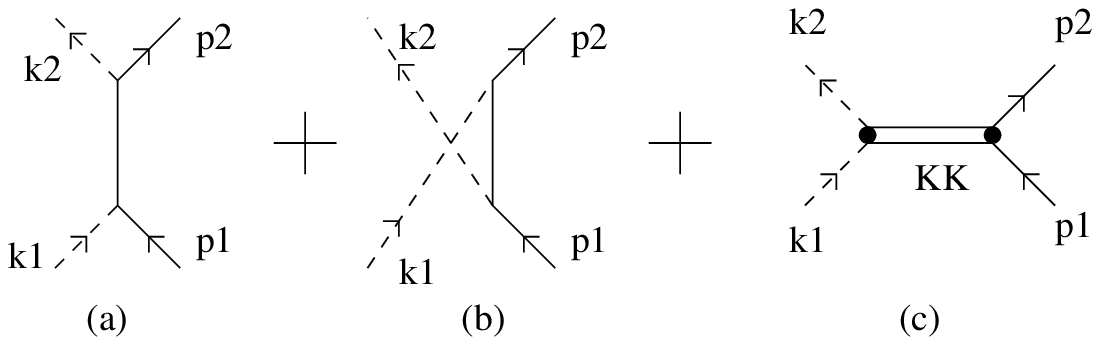}
\end{center}
\caption{Diagrams for high energy Compton scattering, (a) and (b) are the usual 
QED contributions, whereas (c) represents the contribution from exchange of a 
KK-graviton}
\label{fig1}
\end{figure}

\hspace{5pt}\\
\indent The Standard model (SM) has so far been very successful for
explaining most experimental observations relevant to the model.  However, it
has one very ugly feature namely the hierarchy problem, which refers to the
enormous difference in the order of magnitude of the Planck scale
($M_{pl}\sim 10^{19}$GeV) and the weak scale ($\sim 1$TeV).  Recent ideas that
have been proposed to solve the problem introduce extra compact spatial
dimensions.  In the first idea proposed in this direction, given by Arkani-Hamed,
Dimopoulos and Dvali (ADD)\cite{one}, spacetime was a direct product of a four
dimensional Minkowski spacetime and a compact $n$-manifold.  While gravity
was free to propagate through the extra $n$-spatial dimensions as well, SM
fields were restricted to the four-dimensional spacetime.  The Planck-scale
in (4+$n$)-dimensional world could be much much smaller than $M_{pl}$ - even
as low as the TeV scale.  The weakness of gravity for distances $\gg$ compact
dimension arises not because of a fundamental weakness of gravitational
interaction but because of SM-fields being restricted to lie on a brane of
the entire (4+$n$)-dimensional spacetime.\\
\indent Phenomenologically, the ADD-model implies the existence of a large
number of `massive' gravitons coupled with (four-dimensional) Planck strength
to SM-particles, with masses ranging from near zero to some high TeV scale
$M_s$.\\
\indent An alternative scenario considered first by Randall and Sundrum (RS)\cite{two},
the extra dimension is a single $S^1/Z_2$ orbifold with 3-branes residing at
the boundaries of the spacetime.  With suitably tuned brane tensions and a
bulk cosmological constant, RS obtain a non-factorizable metric
\begin{eqnarray}
ds^2 = e^{-2kr_c|\phi|}\eta_{\mu\nu}dx^{\mu}dx^{\nu} - r_c^2d\phi^2
\end{eqnarray}
as a solution of five-dimensional Einstein equations.  Here $\phi\epsilon
[-\pi,\pi]$ is the fifth coordinate with $\phi$ and $-\phi$ identified.  $r_c$
is the radius of the fifth dimension and $k$ is a parameter $\sim M_{pl}$.
Choosing $kr_c \sim 12$, this model also solves the hierarchy problem.
Phenomenologically , the RS-model gives rise to a discrete series of `heavy'
gravitons of increasing mass which couple with SM-particles with weak
interaction strength (referred to now on as KK).\\
\indent Phenomenological consequences of these excited `gravitons' would
naturally be felt at TeV-scale processes.  In situations where the normal SM
contributions are absent or small for some reason the KK-contributions
are obviously the ones that are best suited for testing the theory.
One such vertex is the KK-$\gamma\gamma$ vertex which exists in this theory,
whereas the corresponding $Z\gamma\gamma$ or $H\gamma\gamma$ vertices are
absent \cite{three}.  The existence of the KK-$\gamma\gamma$ vertex would imply a
$t$-channel pole contribution to the Compton-scattering process $\gamma +
e \to \gamma + e$.  The SM-contribution with $s$- and $u$-channel electron
pole will have comparable strength and hence the KK-contribution will show up in the
cross-section of this process.  The feasibility of performing the Compton
process using backscattering laser beams in the NLC has been discussed by
Davoudiasl \cite{four}, who has also calculated the effect of ADD-type
excitations on the process.  Phenomenologically, the RS-model has
differences from the ADD and it will be useful to examine Compton
scattering in the RS-model as well.  In this note we carry out the
analysis not only to see the nature of changes in the cross-section
compared to SM, but also to see if the process at high energies can 
distinguish between the two gravity models.\\
\indent We consider the process
\begin{eqnarray}
\gamma (k_1) + e(p_1) \to \gamma (k_2) + e(p_2) \nonumber
\end{eqnarray}
and neglect the electron mass.  The invariant amplitude for this process is
denoted by $M_{\lambda_1,\lambda_2,\lambda_3,\lambda_4}(s,t,u)$ where $s,t,u$
are the usual Mandelstam variables and $\lambda_1, \lambda_2, \lambda_3,
\lambda_4$ denote respectively the helicities of initial photon, initial
electron, final photon and final electron.  The interaction of the
KK-excitations $h_{\alpha\beta}^{(n)}$ with SM-fields is described by the
interaction Lagrangian \cite{five};
\begin{eqnarray}\label{one}
{\cal L} = -\frac{1}{\Lambda_{\pi}}\cdot T^{\alpha\beta}
h_{\alpha\beta}^{(n)} 
\end{eqnarray}
where $T^{\alpha\beta}$ is the SM-energy momentum tensor.  The constant
$\Lambda_{\pi}$ is related to the parameters of RS metric and the Planck
scale by
\begin{eqnarray}
\Lambda_{\pi} = M_{pl}\cdot e^{-kr_c \pi} .
\end{eqnarray}
For purposes of phenomenological parametrization it is more convenient to
choose the mass of the lowest KK-resonance $m_1$ and
$\left(\frac{k}{M_{pl}}\right)$ as independent parameters with
\begin{eqnarray}
\Lambda_{\pi}^{-1} = \left(\frac{k}{M_{pl}}\right) \cdot
\left(\frac{x_1}{m_1}\right) ,\nonumber
\end{eqnarray}
where $x_1$ is the first zero of $J_1(x)$.  For our numerical evaluation, we
shall fix $m_1$ at a reasonable value of 600GeV and vary
$\left(\frac{k}{M_{pl}}\right)$ over a range of values within the acceptable
limits \cite{five}.\\
\indent With the Lagrangian, equation (\ref{one}), added to the standard
Q.E.D. Lagrangian, the amplitudes for the diagrams of figure (\ref{fig1}) 
work out to be
\begin{eqnarray}
M_{++++}(s,t,u) = M_{----}(s,t,u) & = & \frac{8\pi\alpha\sqrt{s}}{\sqrt{-u}}
- \sum_{i=1}^{12} \frac{2s^{\frac{3}{2}}\sqrt{-u}}{\Lambda_{\pi}^2 (m_i^2
- t)} \\
M_{+-+-}(s,t,u) = M_{-+-+}(s,t,u) & = & \frac{8\pi\alpha\sqrt{-u}}{\sqrt{s}}
+ \sum_{i=1}^{12} \frac{2u\sqrt{-su}}{\Lambda_{\pi}^2 (m_i^2 - t)}
\end{eqnarray}
with all other amplitudes vanishing.  We have retained in the summation
in the right hand side of equation (4) and (5) the first 12 resonaces,
having checked that the terms left out contribute less than 1\%.  In
equations (4) and (5), $m_i$'s are the masses of the KK-resonance of the
spin-2 graviton in the RS model, given by
\begin{eqnarray}
m_i=\left(\frac{k}{M_{pl}}\right) \Lambda_{\pi} \cdot x_i \nonumber
\end{eqnarray}
where the $x_i$'s are the zeroes of the Bessel function $J_1(x)$.  Thus,
for a given $m_1$, which we have taken to be 600 Gev., the $m_i$'s are
simply in proportion to the $x_i$'s. In the limit of vanishing
electron-mass, the contribution of the KK-exchange graphs like the
QED-contribution conserve both electron and photon helicities.\\
\indent The differential cross-section for the two different initial electron
polarizations $j$ ($j=\pm 1$) are then given by
\begin{eqnarray}
\left( \frac{d\sigma}{d\Omega}\right)_j = \frac{1}{64\pi^2s}\cdot \left|
M_{+j+j} \right|^2 .
\end{eqnarray}
The $\gamma$-beams are produced by backscattering  of laser beams off electron
beams; it is thus more appropriate to express the polarized cross-sections in terms
of the helicities of the original laser beam ($p_l$) and the helicity of the
electron beam ($p_e$) from which it was backscattered \cite{four}.  Since the
backscattered photon carries only a fraction $x$ of the energy of the electron
beam, the Mandelstam variables entering the amplitudes $M$ will be
($xs,xt,xu$) instead of ($s,t,u$).  The fraction $x$ theoretically has the
maximum value
\begin{eqnarray}\label{17}
x_{max} = \frac{z}{1+z} \hspace{1em}, \hspace{0.5em} z = \frac{4E_eE_l}{m_e^2}
\end{eqnarray}
where $E_l$ and $E_e$ respectively are the energies of the laser beam and the
electron beam from which it is backscattered.  From equation (7), the value of
$x_{max}$ approaches unity as $z$ tends towards infinity with $E_l$ increasing
indefinitely.  However as $E_l$ increases, $e^+ e^-$ are produced more and more
because of the interaction of the laser beam with the backscattered photons
\cite{six}.  The luminosity of the beam therefore erodes rapidly.  An optimal value
of $z$ is thus finite and is given by \cite{six}
\begin{eqnarray}
z_{opt} = 2 (1+\sqrt{2})
\end{eqnarray}
which is the value we would use in equation (\ref{17}).  For a given value of $x$,
$p_l$, and $p_e$, the photon number density $f(x,p_e,p_l)$ and the average
helicity $\xi_2(x,p_e,p_l)$ are given by \cite{four}:
\begin{eqnarray}
f(x,p_e,p_l) = N^{-1}\left[ \frac{1}{1-x} + (1-x) - 4r(1-r) - r p_l p_e z
(2r-1) (2-x) \right] \\
\xi_2(x,p_e,p_l) = \frac{1}{Nf(x)} \left\{
p_e\left[\frac{x}{1-x}+x(2r-1)^2\right] 
-p_l(2r-1)\left(1-x+\frac{1}{1-x}\right) \right\}
\end{eqnarray}
where $\displaystyle r = \frac{x}{z(1-x)}$ and
\begin{eqnarray}
N & = & \ln (1+z) \cdot \left( 1 - \frac{4}{z} - \frac{8}{z^2} \right) +
\frac{1}{2} + \frac{8}{z} - \frac{1}{2(z+1)^2} \nonumber \\
& & \hspace{2em} + p_e p_l\cdot \left[ \ln
(1+z) \left(1 + \frac{2}{z}\right) - \frac{5}{2} + \frac{1}{1+z} -
\frac{1}{2(z+1)^2} \right]
\end{eqnarray}
\indent For a given $\xi_2$ and a polarization $p_{e_2}$ the electron beam
undergoing Compton scattering, the cross-section can be written in terms of the
projection operators $P_{ij} = \frac{1}{4}(1+ (-1)^i \xi_2).(1+ (-1)^j p_{e_2})$
as
\begin{eqnarray}\label{112}
\frac{d\sigma}{d\Omega} & = & \frac{1}{64\pi^2} \int \frac{f(x)\,dx}{xs} \left[
P_{++}\left|M_{++++}(xs,xt)\right|^2 + P_{+-}\left|M_{+-+-}(xs,xt)\right|^2 +
\right. \nonumber \\
& & \left. \hspace{9em} P_{-+}\left|M_{-+-+}(xs,xt)\right|^2 +
P_{--}\left|M_{----}(xs,xt)\right|^2 \right] \nonumber \\
& = & \frac{1}{128\pi^2} \int \frac{f(x)\,dx}{xs} \left[ \left(\frac{1+\xi_2
p_{e_2}}{2}\right) \left|M_{++++}(xs,xt)\right|^2 + \right. \nonumber \\
& & \left. \hspace{10em} \left(\frac{1-\xi_2 p_{e_2}}{2}\right)
\left|M_{+-+-}(xs,xt)\right|^2 \right]
\end{eqnarray}
where we have made use of parity invariance.  For computational purposes, we
restrict the $x$ integration in equation (\ref{112}) to $x\epsilon [0.1,x_{max}]$ 
and for computing the total cross-sections from equation (\ref{112}), we made 
the cut $\theta_{CM} \epsilon [\frac{\pi}{6},\frac{5\pi}{6}]$ as has been done in
\cite{four}.  Figures 2-4 are our results for the 
unpolarized and polarized cross-sections for a range of values of the parameter
($\frac{k}{M_{pl}}$).  For comparison, the corresponding SM-values and the values 
based on ADD-scenarios as calculated in \cite{four} for n=4 and $M_s=2$ TeV are 
also given.\\
\indent The cross-sections in the present case as a function of $s$ show an
interesting pattern.  At low $s$, the cross-section remains below the SM value,
with the difference increasing with the parameter $k/M_{pl}$.  This is because
in the present case both the SM and the graviton exchange contributions are
real and of opposite signs.  The graviton contribution which increases with
increasing $k/M_{pl}$ thus reduces the amplitude and lowers the cross-section. 
 With increasing energy, the graviton contribution starts dominating and for
$k/M_{pl}$ above 0.05 starts dominating the amplitude from $\sqrt{s} = 1$TeV 
resulting in a monotonic behaviour of the cross-section with the parameter 
$k/M_{pl}$ in that range.  In the case where the graviton enters as a direct
channel resonance \cite{five}, the situation is different.  There the SM
contribution and the graviton contribution add up constructively throughout
resulting in a monotonic dependence of the cross-section with the parameter
$k/M_{pl}$.\\
\indent Our results show that for values of the parameter $k/M_{pl}$ equal to
0.10, the cross-sections are much higher than for SM as for SM with the 
inclusion of ADD grvitons, as $\sqrt{s}$ exceed 1 TeV.  However, for a lower
value of $k/M_{pl}$=0.05, the deviation from SM does not occur till
$\sqrt{s}=1.5$TeV.  For even lower values of $k/M_{pl}$ deviations from SM
would occur at even higher value of $s$.  As discussed in reference 5,
arguments based on heterotic string theory favour $k/M_{pl}=0.01$ and our
results show that for such low values of the parameter, differences with SM
would be insignificant all the way up to $\sqrt{s}=2$ TeV.  This is similar to
the direct observation signal of resonant peaks in the reaction $e^+
e^- \rightarrow\mu^+ \mu^-$\cite{five}, where for $k/M_{pl}=0.01$ the
resonant peak is about ten times the background; for a lower
$k/M_{pl}=0.01$, the peak practically disappears.\\
\indent The polarized cross-section results show that the cross-section for the
$(+,-,+)$ combination of $(p_{e1},p_{l1},p_{l2})$ dominates over the other
combinations in the TeV range.  This is expected since the energy distribution
of the backscattered photons is strongly peaked for the polarizaion combination
$p_{e1},p_{l1}=-1$ \cite{seven}.  This, combined with the helicity
dependence of the amplitudes (12) makes the $(+, -, +)$ cross-sections the
largest one at energies in the TeV range.\\
\indent The signature of the presence of spin-2 graviton exchange is different
in the present case than the one for $e^+ e^- \rightarrow \mu^+ \mu^-$
\cite{five}.  The angular dependence of the cross-section, once the KK-term
starts dominating is peaked in the forward direction unlike the pure SM case,
where the differential cross-section is peaked in the backward direction.  This
feature of course is shared also in the ADD scenario.  However, for high
values of $k/M_{pl}$ where the cross-sections in the RS scenario are much
higher than in the ADD scenario, the pattern of the forward peaking in the
two cases is different.  Wheras in the ADD case, the differential
cross-section in the forward direction to the one at $\theta=\frac{\pi}{2}$is a factor around 3, in the
RS-case, it is over 10.  Hopefully, this feature may help discriminate
between the two models if high energy cross-sections reveal the presence
of spin-2 graviton exchange in the $t$-channel.\\
\indent In terms of significance of the cross-section experimentally, the 
feature of the total cross-section that to begin with at the low energy side,
the cross-section should be lower than the SM value and then should start
increasing, ultimately becoming greater than the SM value should be the most interesting one.  The exact values where the changeover occurs
of course depends
very much on the parameter $k/M_{pl}$.  Formally, we can calculate a $\chi^2$
defined as
\begin{eqnarray}
\chi^2= L \frac{(\sigma(SM+RS)-\sigma(SM))^2}{\sigma(SM)}
\end{eqnarray}
where $L$ is the integrated luminosity over the period of observation and the
$\sigma$'s refer to cross-sections in the present model and in SM.  The value
of $\chi^2$ is of course a function of $s$ and assuming a value of $L = 100 
(fb)^{-1}$, we get the values as listed in table I.  Noting that a 95\%
confidence level implies a value of $\chi^2$ around 2.7, it seems from the
table that it will be possible to see a signature of the RS-graviton resonances
in electron compton scattering in the TeV range if the value of the parameter
is not too low.\\
\indent In conclusion, experimental study of electron Compton scattering in the
TeV range which is feasible with laser backscattering of electron beams in the
NLC would be very useful for revealing the direct coupling to $\gamma\gamma$ of
spin-2 Kaluza-Klein type excitations present in Weak scale Quantum gravity
theories.  This feature will not be shared in other scenarios like supersymmetry
extensions of SM for new Physics at the TeV scale.\\
\indent S.R.Choudhury thanks Prof.B.McKellar and the School of Physics, University 
of Melbourne for their hospitality.

\begin{figure}
\includegraphics[angle=270,width=10cm]{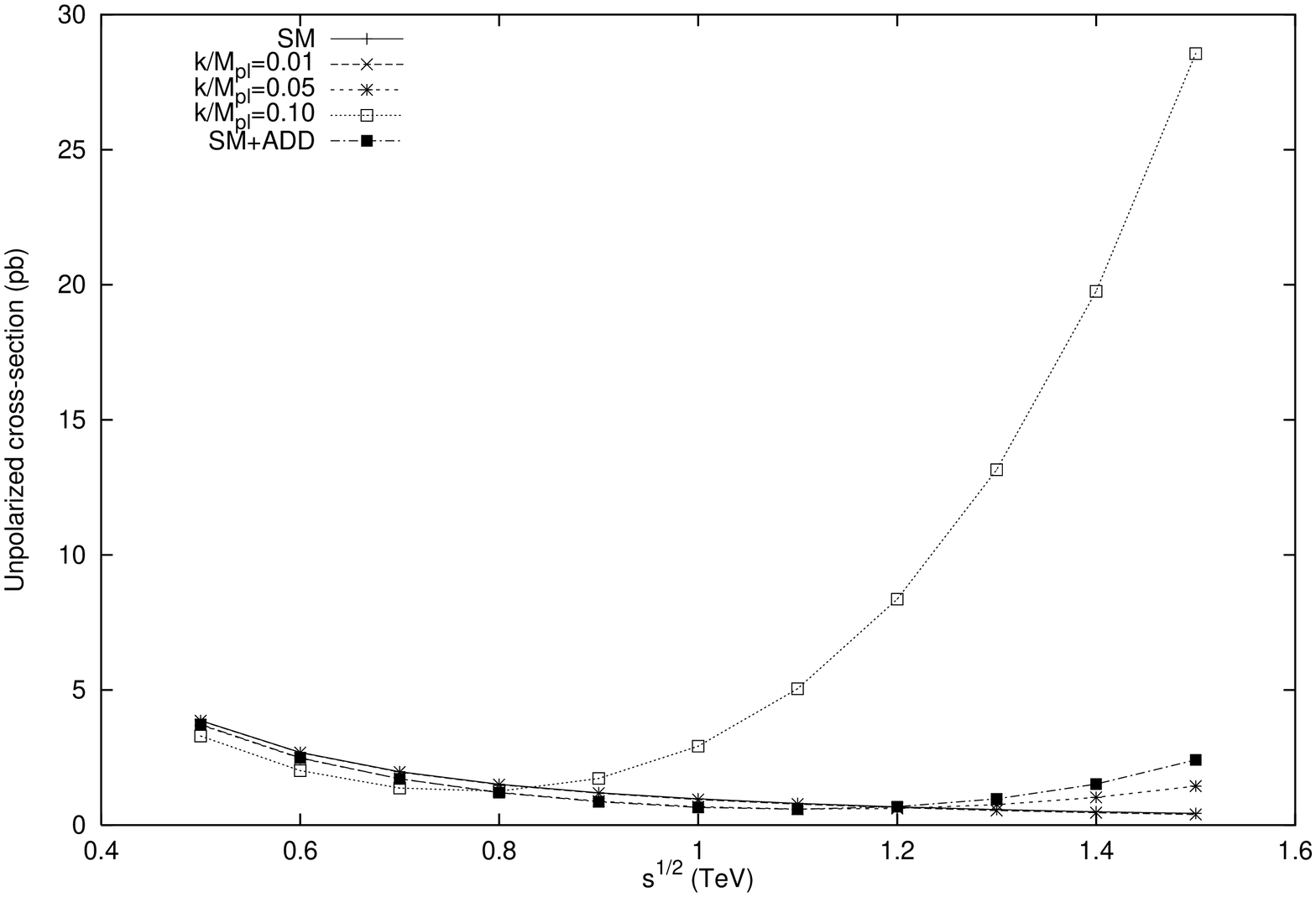}
\caption{Unpolarized cross-sections for $k/M_{pl}$=0.01, 0.05 and 0.10 together
with the SM and SM+ADD values.  The ADD contribution corresponds to the choice
$n$=4, $M_s= 2$TeV, and $\omega=+1$.}
\label{fig2}
\end{figure}

\break

\pagestyle{empty}
\voffset=-0.8in
\begin{figure}
\includegraphics[angle=270,width=10cm]{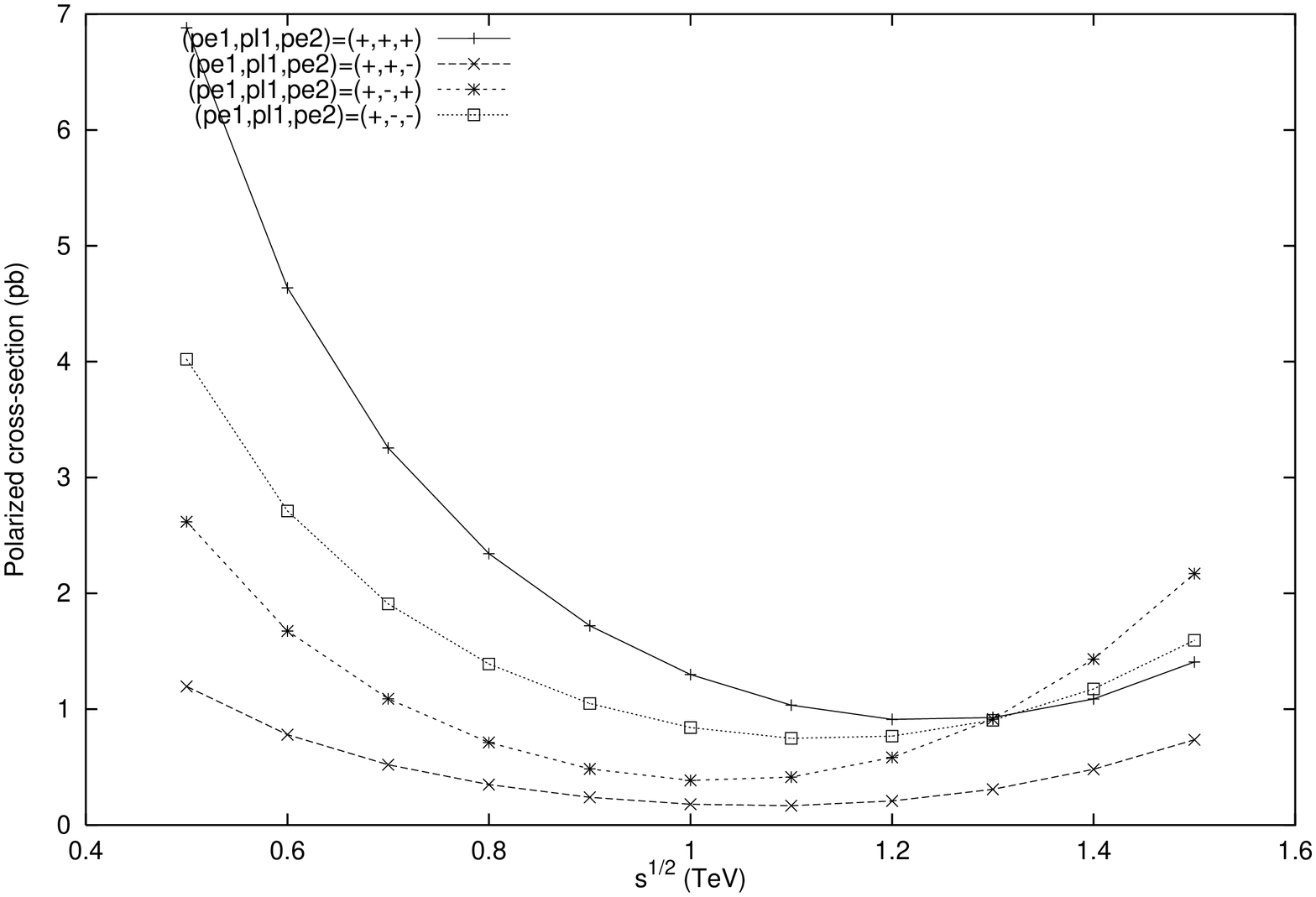}\\
{\small figure 3(a).}\\
\includegraphics[angle=270,width=10cm]{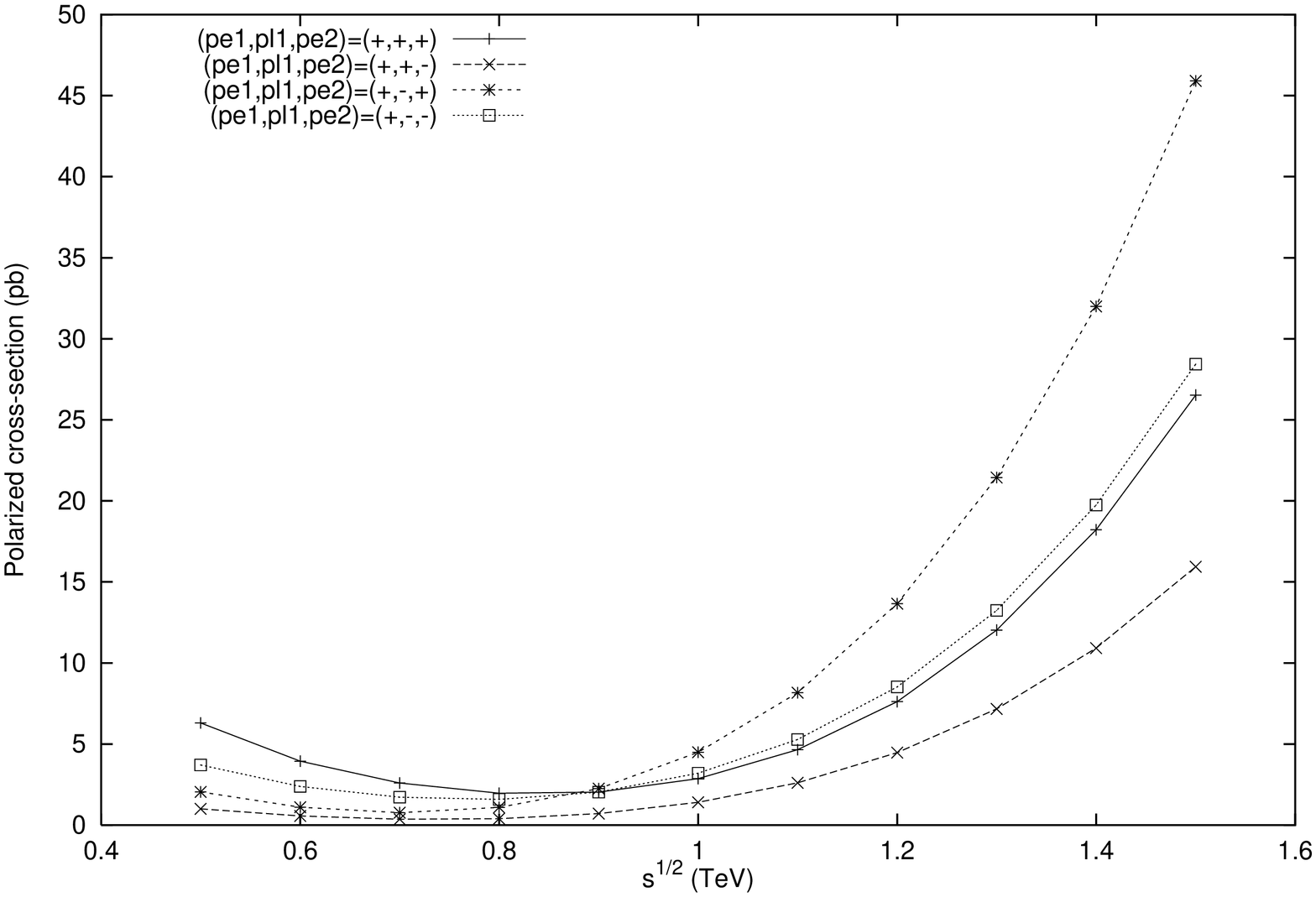}\\
{\small figure 3(b).}\\
\includegraphics[angle=270,width=10cm]{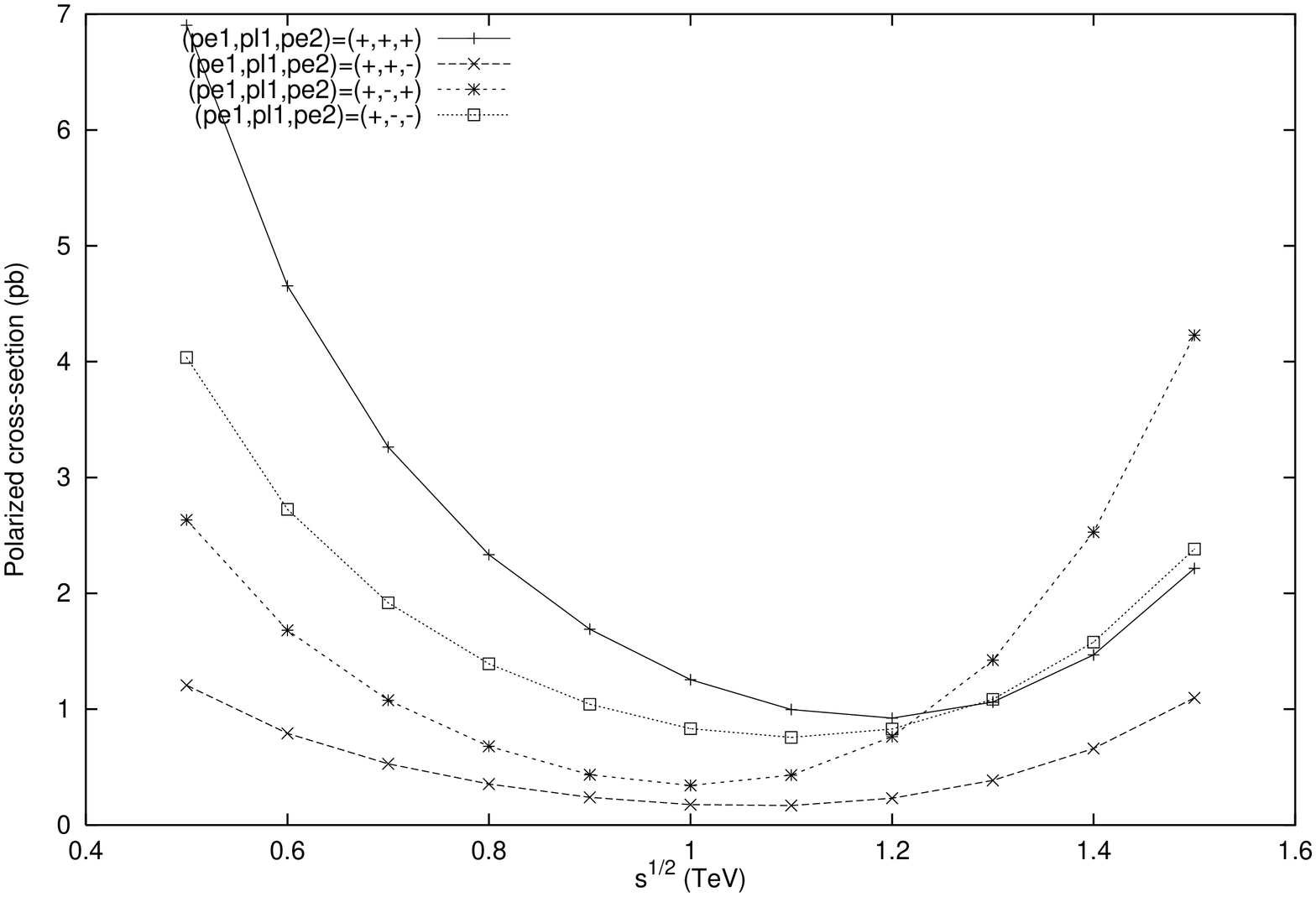}\\
{\small figure 3(c).}
\caption{The total polarized cross-sections as a function of energy:  (a) and
(b) correspond to the RS model with $k/M_{pl}$=0.05 and 0.10 respectively, (c) 
corresponds to SM and SM+ADD value (evaluated with $n$=4, $M_s=2$TeV and 
$\omega=+1$).}
\label{fig3}
\end{figure}

\begin{figure}
\includegraphics[angle=270,width=10cm]{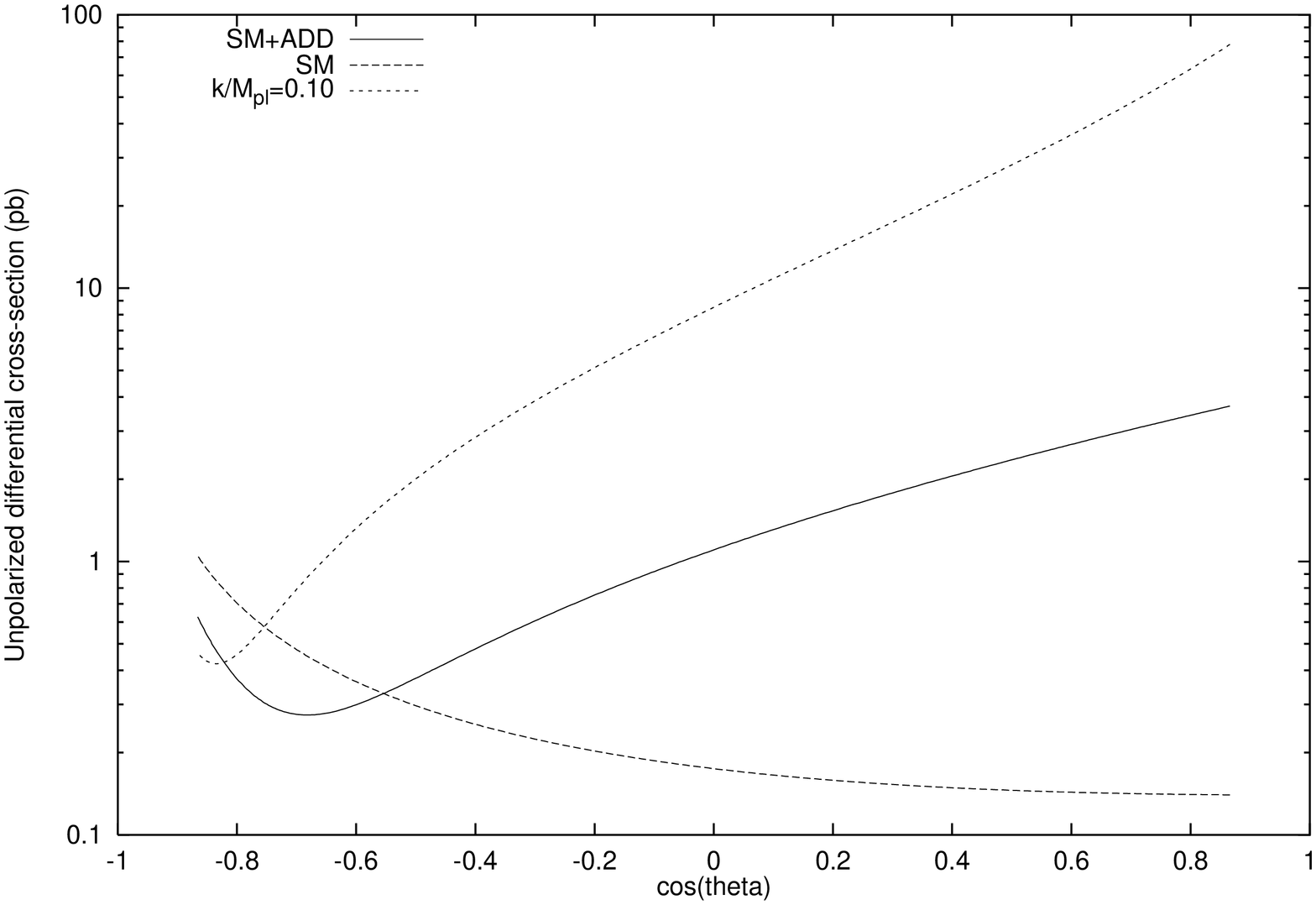}
\\{\small figure 4(a).}\\
\includegraphics[angle=270,width=10cm]{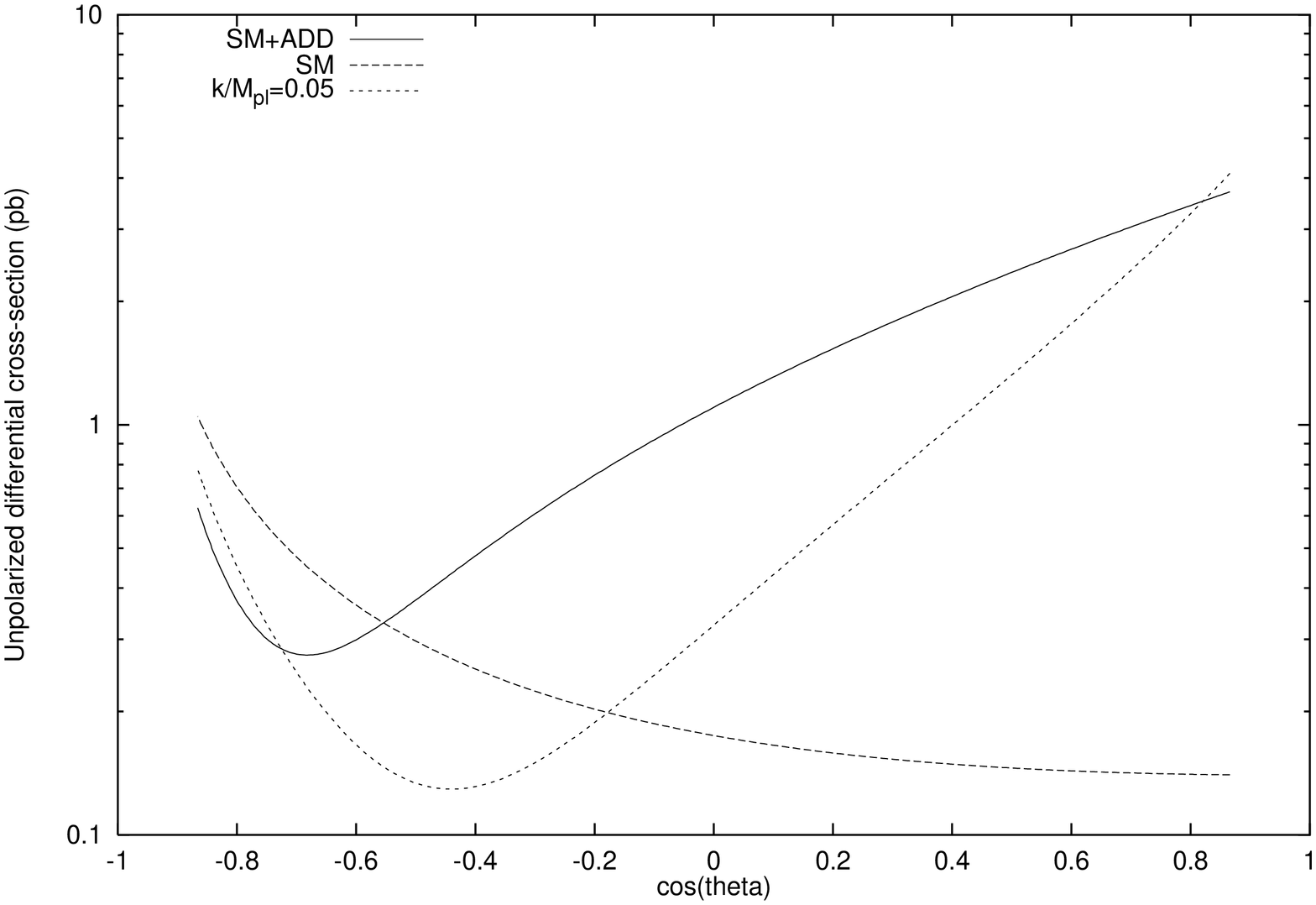}\\
{\small figure 4(b).}
\caption{Unpolarized differential cross-section at $\sqrt{s}=1.5$TeV in the RS 
model for $k/M_{pl}=0.10$ (a) and 0.05 (b).  For comparison, the SM value as
well as value for SM+ADD (evaluated with $n$=4, $M_s = 2$TeV and $\omega=+1$) 
are shown.}
\label{fig4}
\end{figure}

\begin{figure}
\includegraphics[angle=270,width=10cm]{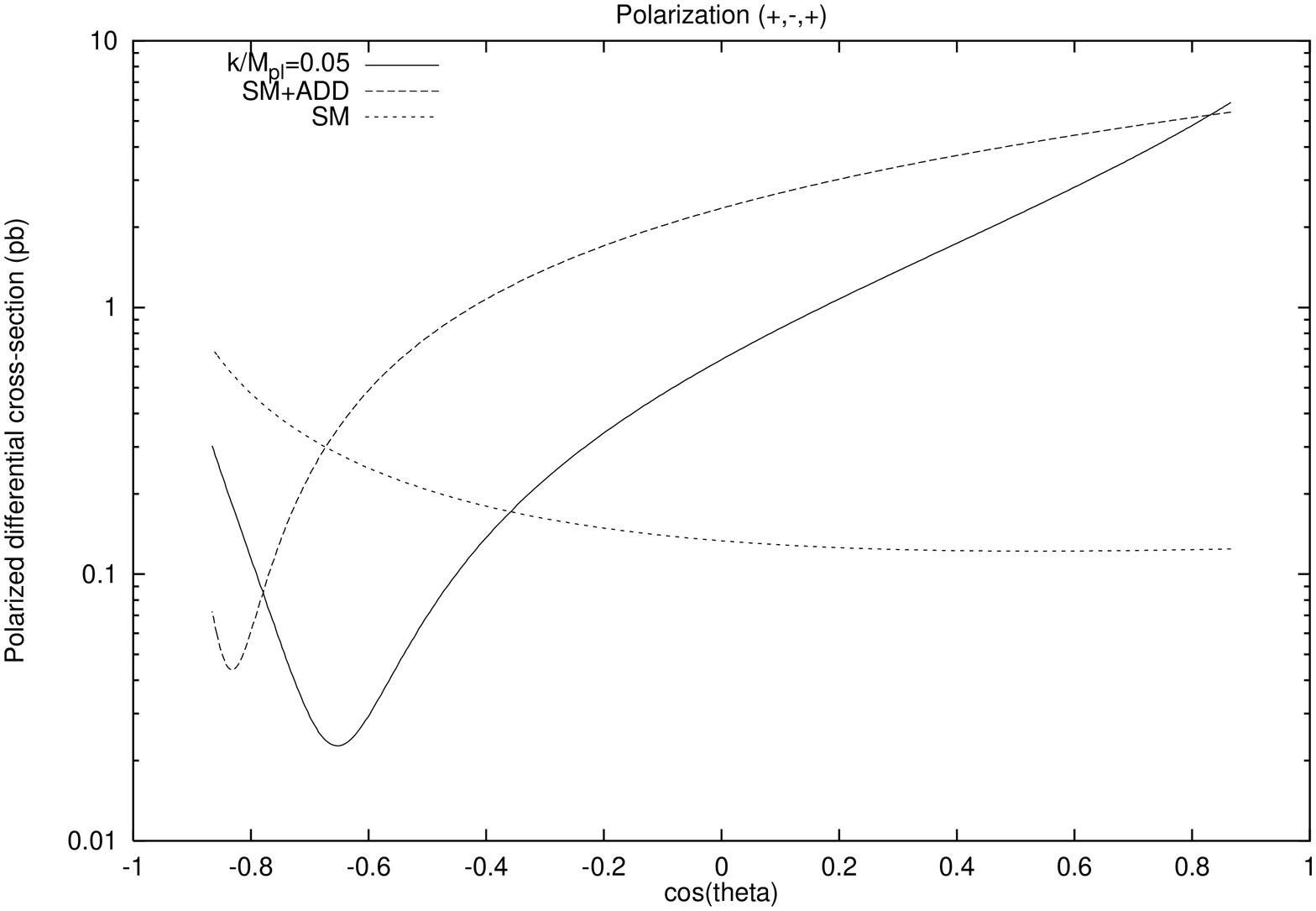}\\
{\small figure 5(a).}\\
\includegraphics[angle=270,width=10cm]{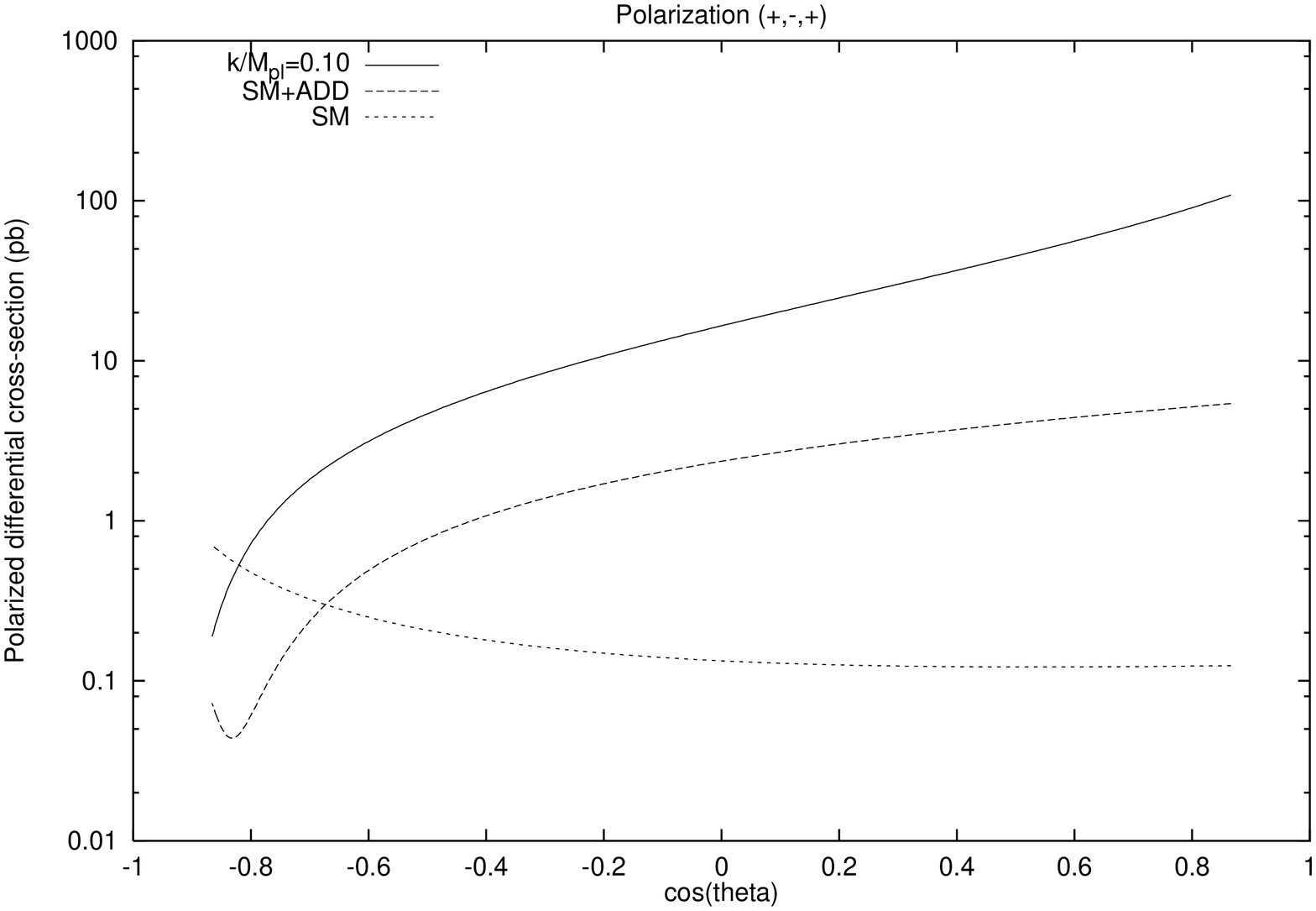}\\
{\small figure 5(b).}
\caption{Polarized differential cross-section for the dominant combination
$p_{e1}, p_{l1}, p_{e2}$  $(+, -, +)$ at $\sqrt{s}=1.5$TeV for $k/M_{pl}$=0.10
(a) and 0.05 (b).  For comparison SM and SM+ADD values (evaluated for
parameters as in figure 4) are also shown.}
\label{fig5}
\end{figure}
\break
\begin{table}[c]
\begin{center}
\begin{tabular}{ccc}
$k/M_{pl}$ & $\sqrt{s}$=0.5 TeV & $\sqrt{s}$=1.0 TeV \\
\hline
0.01 & $10^{-3}$ & 0.04 \\
0.05 & 1.64 & 8.41 \\
0.10 & 16.01 & 326
\end{tabular}
\caption{Table showing $\chi^2$ values for two values of $s$, calculated 
through equation (13) for the cross sections evaluated in the present
calculation including SM and RS-model KK exchange contribution.}
\end{center}
\end{table}

\end{document}